\documentclass[aps,prb,twocolumn,showpacs]{revtex4-1}

\usepackage{amsmath,amssymb}
\usepackage{graphicx}
\usepackage{hyperref}

\newcommand{\itr}{I_\mathrm{tr}}

\newcommand{\pp}{\mathrm{p}}

\begin{document}

\title{Bistability and nonequilibrium transitions in the optically
  polarized system of cavity polaritons under nanosecond-long resonant
  excitation}

\date{\today}

\author{S. S. Gavrilov}
\email{gavr_ss@issp.ac.ru}
\affiliation{Institute of Solid State Physics, RAS, Chernogolovka
  142432, Russia}
\author{A. S. Brichkin}
\affiliation{Institute of Solid State Physics, RAS, Chernogolovka
  142432, Russia}
\author{A. A. Demenev}
\affiliation{Institute of Solid State Physics, RAS, Chernogolovka
  142432, Russia}
\author{A. A. Dorodnyy}
\affiliation{Institute of Solid State Physics, RAS, Chernogolovka
  142432, Russia}
\author{S. I. Novikov}
\affiliation{Institute of Solid State Physics, RAS, Chernogolovka
  142432, Russia}
\author{S. G. Tikhodeev}
\affiliation{A.\,M.\,Prokhorov General Physics Institute, RAS, Moscow
  119991, Russia}
\author{N. A. Gippius}
\affiliation{A.\,M.\,Prokhorov General Physics Institute, RAS, Moscow
  119991, Russia}
\affiliation{LASMEA, UMR 6602 CNRS, Universit\'{e} Blaise Pascal,
  63177 Aubi\`{e}re, France}
\author{V. D. Kulakovskii}
\affiliation{Institute of Solid State Physics, RAS, Chernogolovka
  142432, Russia}

\begin{abstract}
  The polarization dependence of nonequilibrium transitions in a
  multistable cavity-polariton system is studied under a nanosecond
  long resonant optical excitation at the normal and magic angle
  incidences with various polarizations of the pump beam. The temporal
  correlations between the frequency, intensity, and optical
  polarization of the intra-cavity field, which all undergo sharp
  threshold-like changes due to the spin dependent interaction of
  cavity polaritons, are visualized. The observed dynamics cannot be
  reproduced within the conventional semi-classical model based on the
  Gross-Pitaevskii equations. To explain the observed phenomena, it is
  necessary to take into account the unpolarized exciton reservoir
  which brings on additional blueshift of bright excitons, equal in
  the $\sigma^+$ and $\sigma^-$ polarization components. This model
  explains the effect of polarization instability under both pulsed
  and continuous wave resonant excitation conditions, consistently
  with the spin ring pattern formation that has recently been observed
  under Gaussian shaped excitation.
\end{abstract}


\maketitle

\section{Introduction}%
\label{sec:intro}

Cavity polaritons are bound light-matter states that appear due to the
strong coupling of 2D excitons and photons in semiconductor
microcavities.\cite{Weisbuch92,Kavokin-book-03} The optically driven
system of polaritons behaves like a weakly imperfect Bose gas, which
results in a spectacular row of polariton collective phenomena such as
multistability,\cite{Baas04-pra,Gippius07,Paraiso10} parametric
scattering,\cite{Ciuti01,Whittaker01,Gippius04-epl} pattern
formation,\cite{Krizhanovskii10,Sanvitto10} self-organization
effects,\cite{Krizhanovskii08, Demenev08} dynamical Bose-Einstein
condensation.\cite{Kasprzak06} Polariton multistability attracts much
attention as a potential candidate for very fast picosecond range
optical switching on a micron size
scale.\cite{Shelykh08-prl,Liew08-prl-neur} Moreover, the sharp
transitions in intra-cavity field, stemming from the multistability,
can proceed concurrently with inter-mode parametric scattering, which
determines interesting ways of polariton self-organization under
pumping near the inflection point of the lower polariton branch (the
``magic angle'').\cite{Gippius04-epl,Krizhanovskii08,Demenev08}

The polariton bistability has recently been studied in the ``scalar''
approximation neglecting the spin degrees of
freedom.\cite{Baas04-pra,Gippius04-epl} Because of the mutual
interaction of polaritons, their energy effectively depends on the
intra-cavity field magnitude. The repulsion between excitons leads to
the blueshift of polariton energy. When the pump is itself
blue-detuned from the polariton resonance, a strong positive feedback
loop between the resonance energy and the field amplitude is created
in a certain range of system parameters, resulting in sharp jumps in
the intra-cavity field and, hence, in the cavity
transmission. Consequently, the transmission signal can exhibit a
prominent hysteresis in the dependence on continuous-wave (cw) pump
intensity.\cite{Baas04-pra,Gippius04-epl}

The response of the optically excited polariton system becomes more
complicated with allowance made for exciton spin degrees of freedom
(that corresponds to arbitrary optical polarization of the
intra-cavity field). In the general case the system has up to four
stable states under a given cw pump, whereas the actual state of the
system is determined by the history of the excitation
process.\cite{Gippius07,Gavrilov10-en} In critical points, where a
number and/or stability of stationary solutions change, the system can
exhibit sharp jumps in both the amplitude and polarization of the
intra-cavity field. Under a spatially inhomogeneous (e.\,g., Gaussian
shaped) cw excitation the system can also exhibit a nontrivial spatial
distribution of polarization of the luminescence signal, like the
``spin ring'' patterns,\cite{Shelykh08-prl} due to the same underlying
phenomena.

The strong multistability effect predicted in
Ref.~\onlinecite{Gippius07} has recently been observed
experimentally,\cite{Paraiso10} including the sharp jumps in the
cavity transmission under a smooth variation of the pump polarization
degree. The spin ring patterns under Gaussian shaped excitation have
also been reported.\cite{Sarkar10,Adrados10} All these experiments
were carried out under the cw excitation. Thus, one of the still
remaining questions is the dynamical peculiarities of the transitions
in a multistable system, which could only be traced using the
time-resolved techniques. Particularly, the characteristic switching
times between different stability branches call for the experimental
study, for they can have a crucial impact upon the practical
implementations.

Another question concerns the theoretical approach allowing to
describe the observed multistability effect in
microcavities. Traditionally, the multistability is considered in
terms of the self-acting classical fields corresponding to
macro-occupied coherent polariton modes which appear under a coherent
resonant
excitation.\cite{Baas04-pra,Gippius07,Shelykh08-prl,Liew08-prl-neur}
Although such an approach is supposed to be sufficient to describe the
bistability in a circularly polarized system,\cite{Baas04-pra} it
gives wrong predictions for the general case of elliptically polarized
excitation.\cite{Demenev10} Most probably, the incoherent states of
the exciton reservoir which are inevitably excited in optical
experiments have a substantial impact on the decay rates and energies
of polaritons. For instance, the nonlinear decay of polaritons with
different circular polarizations was taken into consideration in order
to explain the experimental data in
Ref.~\onlinecite{Paraiso10}. Further, the nonequilibrium transitions
reported in Ref.~\onlinecite{Sarkar10} can only be reproduced in
calculations taking into account the reservoir induced shifts of the
polariton energy.

In the present work we report the experimental study of nonequilibrium
transitions in the multistable cavity polariton system. Unlike in
recent Refs.~\onlinecite{Paraiso10,Sarkar10,Adrados10}, we have
studied the optically polarized system under a pulsed nanosecond-long
excitation, which allowed us to trace the time-resolved dynamics of
the intra-cavity field. The shifts in polariton energy were reflected
by temporal variations in the transmission energy spectrum. Thus, the
employed technique is capable of visualizing the temporal correlations
between the resonance energy and intensity of the intra-cavity field.

We discuss in detail the time dependence of the transmission signal
polarization, below as well as above the threshold, for several
polarizations of the pump beam. The observed polarization behavior
cannot be reproduced within a semi-classical model based on the
Gross-Pitaevskii equations considered in
Refs.~\onlinecite{Gippius07,Shelykh08-prl,Gavrilov10-en}.  The
experimental results allowed us to develop a phenomenological model to
describe a multistable polariton system with a proper regard to the
exciton reservoir. The reservoir excitons shift the polariton energy
that, in turn, influences the thresholds of nonequilibrium
transitions. Both the linear and nonlinear mechanisms of exciton
scattering into the reservoir are found to be significant within a
sub-nanosecond time scale. While the nonlinear decay of
cross-circularly polarized excitons leads to the levelling of the
$\sigma^+$ and $\sigma^-$ jump points (which has also been found in
Refs.~\onlinecite{Paraiso10,Sarkar10} in the cw pump regime), the
linear decay mechanism leads to the temporal delay of the jumps with
respect to the pump intensity peak.

Further, we apply the developed approach to simulate the parametric
scattering under pumping at the magic angle. As we show, this model
allows to explain self-consistently the temporal dependences of
polarizations of both the driven mode and the scattering signal which
appears at the polariton branch bottom.

The paper is organized as follows. In Sec.~\ref{sec:theory} the
available models of the multistability effect are considered and
compared with each other in view of the recent experimental results of
Refs.~\onlinecite{Demenev10,Paraiso10,Sarkar10}.
Sec.~\ref{experimental} references the experimental setup.
Sec.~\ref{sec:experiment} contains the experimental results and
compares them with the calculations performed in the framework of the
suggested model, for the cases of pumping at normal incidence
(Sec.~\ref{sec:normal}) and at the magic angle (Sec.~\ref{sec:magic}),
with a brief reference to some of the still unresolved issues
(Sec.~\ref{sec:discussion}). The results are summarized in Conclusion
(Sec.~\ref{sec:conclusions}).

\section{Theoretical models}%
\label{sec:theory}

The multistability effects in microcavities were recently considered
in the framework of Gross-Pitaevskii equations written for the
macro-occupied polariton modes.\cite{Gippius07,Shelykh08-prl} These
equations can also be written for the strongly coupled exciton and
photon fields ($\mathcal{P}$ and $\mathcal{E}$ respectively) in the
cavity active layer:\cite{Gavrilov10-en}
\begin{align}
  i \dot{\mathcal{E}}_+ &= (\omega_\mathrm{c} - i\gamma_\mathrm{c}) \,
  \mathcal{E}_+ + \alpha \mathcal{F}_+ + \beta \mathcal{P}_+, \label{ef_p} \\
  i \dot{\mathcal{E}}_- &= (\omega_\mathrm{c} - i\gamma_\mathrm{c}) \,
  \mathcal{E}_- + \alpha \mathcal{F}_- + \beta \mathcal{P}_-, \label{ef_m} \\
  i \dot{\mathcal{P}}_+ &= \bigl( \omega_\mathrm{x} + V_1
  |\mathcal{P}_+|^2 + V_2
  |\mathcal{P}_-|^2 - i \gamma_\mathrm{x} \bigr) \mathcal{P}_+ + A \mathcal{E}_+, \label{xp_p} \\
  i \dot{\mathcal{P}}_- &= \bigl( \omega_\mathrm{x} + V_2
  |\mathcal{P}_+|^2 + V_1 |\mathcal{P}_-|^2 - i \gamma_\mathrm{x}
  \bigr) \mathcal{P}_- + A \mathcal{E}_- \label{xp_m}
\end{align}
(in the $\sigma^\pm$ basis). Here, $\mathcal{F}$ stands for the
incident electric field that is usually treated as a plane wave:
$\mathcal{F}_\pm \propto e^{-i\omega_\mathrm{p}t}$;
$\omega_\mathrm{c,x}$ and $\gamma_\mathrm{c,x}$ are the
eigenfrequencies and decay rates of the intra-cavity photon and
exciton modes; $\alpha$ and $\beta$ are the cavity response
coefficients, and $A$ is exciton polarizability (so that $2
\sqrt{A\beta}$ equals Rabi splitting); $V_{1,2}$ are the matrix
elements of the interaction between excitons with same ($V_1$) and
opposite ($V_2$) circular polarizations.

For simplicity, Eqs.~(\ref{ef_p}--\ref{xp_m}) are written for only the
driven polariton mode with zero quasi-momentum, although they can be
easily generalized to the case of essentially many-mode
system.\cite{Gavrilov10-en} In the latter case, the usage of the
exciton-photon basis (instead of the simplified polariton basis)
brings the advantage of taking account of the TE/TM splitting effects
as well as the angular dependence of the polariton interaction
constants. As will be shown below, it is also convenient for taking
account of the incoherent exciton reservoir.

The multistability conditions are defined by the relation between
exciton-exciton interaction constants, $V_1$ and $V_2$. According to
the study of polariton spin quantum beats in a GaAs
cavity,\cite{Renucci05} the cross-circularly polarized excitons
exhibit attraction ($V_2 < 0$) that is weaker than the repulsive
interaction of co-circular excitons ($V_1 > 0$ and $|V_2| < V_1$). The
fact that $V_2$ is negative is also confirmed by the rotation of
polariton polarization direction during the process of parametric
scattering.\cite{Krizhanovskii06-prb} In the general case the
processes of exciton-exciton interaction are mediated by biexciton and
dark exciton states,\cite{Inoue00,Wouters07,Schumacher07} hence the
relation between the interaction constants of \emph{polaritons} can be
dependent on the exciton-photon detuning.\cite{Vladimirova10} More
precisely, one should also consider the saturation of excitonic
transitions as an additional nonlinear mechanism affecting the
blueshift of polaritons under a sufficiently strong
excitation.\cite{Gonokami97}

If $V_1 > 0$ and $V_2 \lesssim 0$ (as assumed in
Refs.~\onlinecite{Gippius07,Shelykh08-prl,Gavrilov10-en} basing on the
experimental estimations of Ref.~\onlinecite{Renucci05}), then the
$\sigma^\pm$-components of the intra-cavity field are almost
uncoupled. If $V_2 < 0$, then the threshold intensity $W =
|\mathcal{F}_+|^2 + |\mathcal{F}_-|^2$ of linearly polarized pump is
at least two times larger than that of circularly polarized pump:
$W_\mathrm{thr}^\mathrm{(lin)} \gtrsim 2
W_\mathrm{thr}^\mathrm{(circ)}$.

Contrary to the above predictions, $W_\mathrm{thr}^\mathrm{(lin)}$ has
been found to be slightly less than $W_\mathrm{thr}^\mathrm{(circ)}$
in the experiments using cw as well as time-resolved pumping
techniques.\cite{Demenev10, Paraiso10, Sarkar10}$^,$\footnote{The
  experiments of Ref.~\onlinecite{Demenev10} were performed under the
  time-resolved nanosecond long excitation at the magic
  angle. Further, the relation $W_\mathrm{thr}^\mathrm{(lin)} <
  W_\mathrm{thr}^\mathrm{(circ)}$ was observed under normal incidence
  cw excitation in the 3\,$\mu$m wide cavity mesa~\cite{Paraiso10} and
  (under the same cw pump conditions) in ordinary (spatially uniform)
  microcavity structure.\cite{Sarkar10} In the latter case, the
  negative sign of $V_2$ in a studied sample was proven by the
  observed rotation of the scattering signal polarization. In the case
  of weakly circularly polarized pump, the jumps in both $\sigma^+$
  and $\sigma^-$ modes were found to occur
  simultaneously\cite{Paraiso10,Sarkar10} in the sense of cw
  excitation conditions with ``infinitesimally slow'' variations of
  pump intensity.  In view of the Gross-Pitaevskii
  equations~(\ref{ef_p})--(\ref{xp_m}), this looks like a
  manifestation of a large repulsion between polaritons with opposite
  pseudospins.  At the same time, in a case when the pump energy is
  less than a half of biexciton energy, one expects an attractive
  interaction between excitons with different
  pseudospins,\cite{Wouters07} so they cannot provide blueshift for
  each other; it agrees with the observed indications of the negative
  sign of $V_2$.}%
In order to explain it, two theoretical approaches have been
proposed. The first one\cite{Paraiso10,CerdaMendez10} presumes a
repulsive interaction of polaritons with opposite pseudospins
($\sigma^\pm$ polarizations) in the framework of Gross-Pitaevskii
equations of type (\ref{ef_p})--(\ref{xp_m}).  The second
approach\cite{Sarkar10,Gavrilov10-jetpl-en} considers the additional
blueshift resultant from an exciton reservoir, which leads to
$W_\mathrm{thr}^\mathrm{(lin)} < W_\mathrm{thr}^\mathrm{(circ)}$
without the necessity for a repulsive interaction between
cross-circularly polarized excitons. (A somewhat similar model has
previously been proposed in Ref.~\onlinecite{Wouters07-prl} to
describe the interaction of a ``spinless'' polariton condensate
with a reservoir.)

In the present work we explore the second model accounting for the
transitions of the optically driven excitons into the incoherent
(reservoir) state in which the overall pseudospin is relaxed, so that
the reservoir provides equal blueshifts for both polarization
components of the coherent state.\cite{Sarkar10,Gavrilov10-jetpl-en}
Those transitions are introduced phenomenologically as non-radiative
decays of bright excitons accorded with the rate of occupation of the
optically inactive reservoir. The model for the intra-cavity electric
field $\mathcal{E}$ and exciton polarization $\mathcal{P}$ coupled
with the integral population of the reservoir $\mathcal{N}$ can be
written as
\begin{align}
  i \dot{\mathcal{E}}_+ &= (\omega_\mathrm{c} - i\gamma_\mathrm{c}) \,
  \mathcal{E}_+ + \alpha \mathcal{F}_+ + \beta \mathcal{P}_+, \label{ef_p_r} \\
  i \dot{\mathcal{E}}_- &= (\omega_\mathrm{c} - i\gamma_\mathrm{c}) \,
  \mathcal{E}_- + \alpha \mathcal{F}_- + \beta \mathcal{P}_-, \label{ef_m_r} \\
  i \dot{\mathcal{P}}_+ &= \bigl[ \omega_\mathrm{x} + V_1
  |\mathcal{P}_+|^2 + V_2
  |\mathcal{P}_-|^2 + (V_1 + V_2) \, \mathcal{N}/2 - {} \nonumber \\
  &i \left( \gamma_\mathrm{x} + \gamma_\mathrm{xr} + V_\mathrm{r}
    |\mathcal{P}_-|^2 \right) \bigr]
  \mathcal{P}_+ + A \mathcal{E}_+, \label{xp_p_r} \\
  i \dot{\mathcal{P}}_- &= \bigl[ \omega_\mathrm{x} + V_1
  |\mathcal{P}_-|^2 + V_2
  |\mathcal{P}_+|^2 + (V_1 + V_2) \, \mathcal{N}/2 - {} \nonumber\\
  &i \left( \gamma_\mathrm{x} + \gamma_\mathrm{xr} + V_\mathrm{r}
    |\mathcal{P}_+|^2 \right) \bigr]
  \mathcal{P}_- + A \mathcal{E}_-, \label{xp_m_r} \\
  \dot{\mathcal{N}}~{} &= -\gamma_\mathrm{r} \mathcal{N} + 2
  \gamma_\mathrm{xr} \left( |\mathcal{P}_+|^2 + |\mathcal{P}_-|^2
  \right) + {} \nonumber\\ &4 V_\mathrm{r} |\mathcal{P}_+|^2
  |\mathcal{P}_-|^2. \label{res}
\end{align}
This model is a generalization of Eqs.~(\ref{ef_p})--(\ref{xp_m}).
Here, $\gamma_\mathrm{xr}$ is an additional decay rate of excitons
that corresponds to the linear mechanism of light absorption,
providing the term $2 \gamma_\mathrm{xr} (|\mathcal{P}_+|^2 +
|\mathcal{P}_-|^2)$ in Eq.~(\ref{res}); $V_\mathrm{r}$ stands for the
rate of nonlinear interaction between $\mathcal{P}_\pm$ that provides
an additional occupation of the reservoir ($4 V_\mathrm{r}
|\mathcal{P}_+|^2 |\mathcal{P}_-|^2$ per unit time) due to the mixture
of excitons with opposite polarizations; $\gamma_\mathrm{r}$ stands
for the reservoir own decay rate.\footnote{For the sake of simplicity,
  only the nonlinear mechanism of reservoir population
  ($\gamma_\mathrm{xr} = 0$ and $V_\mathrm{r} \neq 0$) has been
  considered in Ref.~\onlinecite{Sarkar10}. In such a case, the
  reservoir induced blueshifts are determined mainly by the ratio
  between reservoir's decay rate and the coupling constant
  $V_\mathrm{r}$. For instance, the calculations performed with
  different $\gamma_\mathrm{r}$ but constant $\gamma_\mathrm{r} /
  V_\mathrm{r}$ and a sufficiently slow variation of the pump power
  lead to exactly same results for the cases of circular and linear
  pump polarizations. In this connection, the experiments carried out
  with time resolution using the nanosecond long pump pulses
  (described in Sec.~\ref{sec:experiment}) are capable of clarifying
  the characteristic lifetimes of the reservoir states, as well as the
  relation between linear and nonlinear mechanisms of polaritons'
  scattering into the incoherent exciton reservoir. For instance, the
  former mechanism is expected to play a role irrespectively of pump
  polarization, affecting the dynamics of polariton energy and the
  time moments when the nonequilibrium transitions occur.}
Microscopically, the nonlinear absorption of cross-circularly
polarized excitons is resultant from the scattering of a pair of
bright excitons with opposite spins ($J_z = -1$ and $J_z = +1$) into
dark excitons ($J_z = -2$ and $J_z = +2$), which is closely related to
biexciton creation (see Ref.~\onlinecite{Inoue00} for details). We do
not consider the reverse transitions of incoherent (reservoir)
excitons into the driven mode; this approximation is valid as long as
the occupation of the reservoir states is small enough. Under resonant
pumping at $\mathbf k=0$ the energy mismatch of the pump with respect
to free exciton makes the filling of reservoir quite ineffective, but
it is still possible due to energy level broadening and non-zero
temperature. The reservoir can accumulate excitation due to its
comparatively long lifetime (see Sec.~\ref{sec:experiment}).

Equations~(\ref{ef_p_r})--(\ref{res}) can easily be generalized to the
case of many-mode system, with exactly the same nonlinear interaction
terms (which are ``local'' in the real space), and a linear part being
in the $k$-space the same as in Ref.~\onlinecite{Gavrilov10-en}. In
the many-mode calculations represented below in Figs.~\ref{fig:2},
\ref{fig:3}, and \ref{fig:11}--\ref{fig:13}, the cavity dispersion
$E_\mathrm{c}^\mathrm{TE, TM}(k)$ and response coefficients
$\alpha_\mathrm{TE, TM}(k)$, $\beta_\mathrm{TE, TM}(k)$ for TE and TM
cavity modes are properly taken into account using the transfer matrix
technique.\cite{Tikhodeev02}

\begin{figure}[t]
  \centering
  \includegraphics[width=0.99\linewidth]{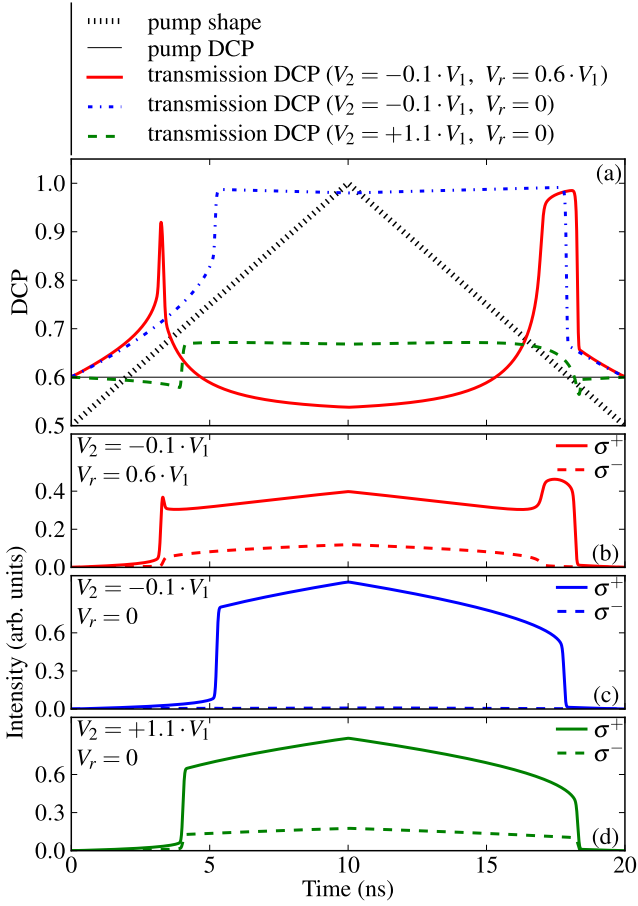}
  \caption{(a) The calculated time dependence of the cavity
    transmission polarization under a slowly varying pump
    (transverse-dashed triangular line) with elliptical
    polarization. The degree of circular polarization (DCP) of pump is
    $\rho_\mathrm{p} = 0.6$ (thin solid line). Thick solid line shows
    the data calculated using Eqs.~(\ref{ef_p_r})--(\ref{res}) with
    $V_2 / V_1 = -0.1$, $V_\mathrm{r} / (\gamma_\mathrm{r} V_1) =
    6\,\mathrm{meV}^{-1}$, and $\gamma_\mathrm{xr} = 0$ (explicitly as
    in Ref.~\onlinecite{Sarkar10}). Dash-and-dot line represents the
    same dependence calculated using Eqs.~(\ref{ef_p})--(\ref{xp_m})
    with $V_2 / V_1 = -0.1$ (this ratio was used in
    Refs.~\onlinecite{Gippius07,Shelykh08-prl}). Dash line corresponds
    to the case of $V_2 / V_1 = +1.1$. The calculated data is
    time-averaged for each temporal point over 100 ps intervals, in
    order to eliminate the fast transitional effects which are not
    observable under the cw pump conditions. (b-d) The corresponding
    time dependences of the intra-cavity field intensity in
    $\sigma^\pm$ polarization components.  }
  \label{fig:1}
\end{figure}
\begin{figure}
  \centering
  \includegraphics[width=1.0\linewidth]{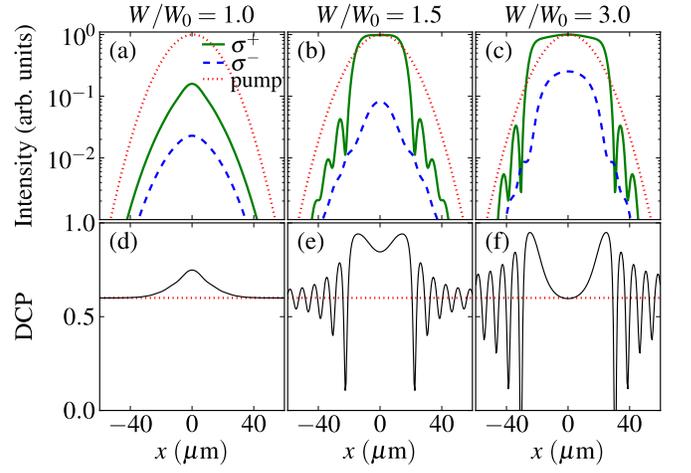}
  \caption{The calculated ``spin ring'' pattern formation in the
    framework of many-mode model Eqs.~(\ref{ef_p_r})--(\ref{res}) with
    $V_\mathrm{r} = 0.6 \cdot V_1$, $V_2 = -0.1 \cdot V_1$, and
    $\gamma_\mathrm{xr} = 0$ (explicitly as in
    Ref.~\onlinecite{Sarkar10}). The pump DCP is $+0.6$ (as in
    Fig.~\ref{fig:1}). Panels \textit{a--e} show the spatial
    distribution of intensity of the intra-cavity field in $\sigma^+$
    and $\sigma^-$ polarization components (solid and dashed lines,
    respectively), for several values of pump power~$W$; the pump
    distribution profile is shown by dotted lines. $W_0$ roughly
    corresponds to the threshold excitation density. Panels
    \textit{d--f} represent the corresponding spatial distribution of
    the intra-cavity field DCP.}
  \label{fig:2}
\end{figure}
\begin{figure}
  \centering
  \includegraphics[width=1.0\linewidth]{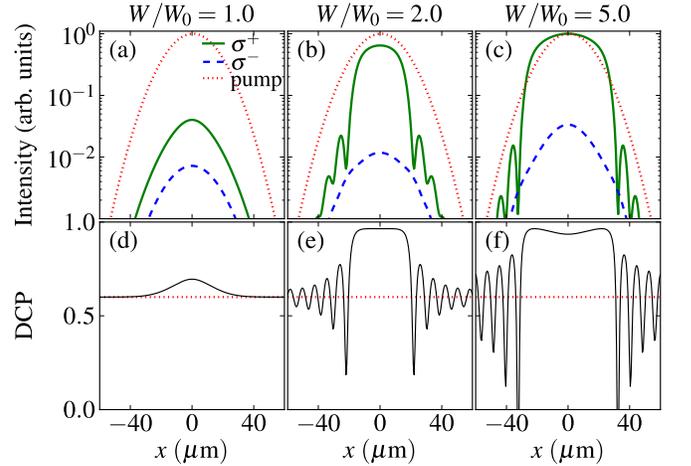}
  \caption{The calculated spatial distribution of the intra-cavity
    field for $V_\mathrm{r} = \gamma_\mathrm{xr} = 0$ and $V_2 = -0.1
    \cdot V_1$. All other parameters coincide with those of
    Fig.~\ref{fig:2}. Unlike Fig.~\ref{fig:2}, almost no ring
    structure appears in the cavity transmission DCP even at $W/W_0 =
    5$.}
  \label{fig:3}
\end{figure}

The two models, Eqs.~(\ref{ef_p})--(\ref{xp_m}) and
(\ref{ef_p_r})--(\ref{res}), can give similar predictions for the
cases of circular and linear polarizations of excitation. In
particular, the key experimental effect waiting for explanation---that
the threshold of linearly polarized pump is smaller
($W_\mathrm{thr}^\mathrm{(lin)} \lesssim
W_\mathrm{thr}^\mathrm{(circ)}$)---can be reproduced within the
Eqs.~(\ref{ef_p})--(\ref{xp_m}) with $V_2 \gtrsim V_1$. Nonetheless,
the transitional dynamics under the intermediate pump polarizations $0
< \rho_\mathrm{p} < 1$ are completely different in the two systems,
see an example in Fig.~\ref{fig:1}.

If $V_2 < 0$ and $V_\mathrm{r} > 0$ (Fig.~\ref{fig:1}\textit{a}, solid
line, and Fig.~\ref{fig:1}\textit{b}), the cavity transmission
polarization increases with pump in the sub-threshold area, due to the
dominant blueshift of the leading polarization component
($\rho_\mathrm{p} = +0.6$). On reaching the threshold, the leading
component jumps up to the high-energy state, which causes a step-like
increase in reservoir population and the blueshift of $\sigma^-$
mode. Consequently, the minor mode also enters the above-threshold
area, and the degree of circular polarization (DCP) of transmission
($\rho_\mathrm{tr}$) decreases down to the values smaller than the
pump DCP. The same effect results in formation of the spin ring
patterns under the Gaussian shaped
excitation\cite{Shelykh08-prl,Sarkar10,Adrados10} in a close vicinity
of the threshold pump power (Fig.~\ref{fig:2}). As a matter of fact,
the ring shaped pattern of DCP indicates that both polarization
components reach the high-energy state at the center of the pump spot
where, therefore, the transmission DCP is minimal. The spatial
distribution of $\rho_\mathrm{tr}$ (Fig.~\ref{fig:2}\textit{d--f})
exhibits a well pronounced minimum at $x=0$ already when the threshold
is surpassed by 1.5 times (Fig.~\ref{fig:2}\textit{e}), in agreement
with the experimental observations (see Fig.\,3 in
Ref.\,\onlinecite{Sarkar10}). This markedly differs from a prediction
of the conventional theory based on Gross-Pitaevskii
equations\cite{Gippius07,Shelykh08-prl,Gavrilov10-en} with $|V_2| \ll
V_1$ (see Fig.~\ref{fig:3}) which does not exhibit the rings even at
$W/W_\mathrm{thr} \approx 5$ for $\rho_\mathrm{p} = 0.6$. The temporal
dynamics for $|V_2| \ll V_1$ is shown in Fig.~\ref{fig:1}\textit{a}
(dash-and-dot line) and in Fig.~\ref{fig:1}\textit{c}.

In the opposite case, when $V_2 > V_1$ and the system is ``purely
coherent'' (Fig.~\ref{fig:1}\textit{a}, dash line, and
Fig.~\ref{fig:1}\textit{d}), which also corresponds to the right
relation $W_\mathrm{thr}^\mathrm{(lin)} <
W_\mathrm{thr}^\mathrm{(circ)}$, the polarization dynamics is
completely different. First, the transmission DCP decreases in the
sub-threshold area, since an increase in the ``leading'' $\sigma^+$
component provides a larger blueshift for $\sigma^-$, so that the
difference in $\sigma^\pm$ intensities weakens with increasing
pump. On reaching the threshold, $\rho_\mathrm{tr}$ shows a minor
step-like increase followed by the very weak changes with further
increasing pump power. No prerequisites for the spin ring formation
are satisfied in this case.

Thus, the observed phenomena---the disparity of thresholds,
$W_\mathrm{thr}^\mathrm{(lin)} \lesssim
W_\mathrm{thr}^\mathrm{(circ)}$, together with the field polarization
dynamics---cannot be reproduced within Eqs.~(\ref{ef_p})--(\ref{xp_m})
using only two constants $V_{1,2}$ of exciton-exciton interaction even
if a single polariton mode at $\mathbf{k} = 0$ is considered. The
experiments revealing both the energy (blueshift) and polarization
dynamics of the driven polariton mode under the nanosecond-long
excitation pulses are described below in Sec.~\ref{sec:experiment}.

\section{Experiment}%
\label{experimental}

The microcavity structure grown by a metal-organic vapor-phase epitaxy
technique has top (bottom) Bragg reflectors composed of 17 (20)
$\lambda/4$ Al$_{0.13}$Ga$_{0.87}$As/AlAs layers. The $3\lambda/2$
GaAs cavity contains six $10$\,nm thick In$_{0.06}$Ga$_{0.94}$As/GaAs
quantum wells. The Rabi splitting is about $6$\,meV. A gradual
variation of the active layer thickness along the sample provides a
change in the photon mode energy $E_\mathrm{c} =
\hbar\omega_\mathrm{c}$ and, accordingly, in the detuning $\Delta$
between the exciton $E_\mathrm{x}(k\,{=}0\,)$ and photon
$E_\mathrm{c}(k\,{=}\,0)$ mode energies. Experiments are carried out
in several regions of the same sample with $\Delta \approx 0$
(Sec.\,~\ref{sec:normal}) and $\Delta \approx -1.5\,\mathrm{meV}$
(Sec.\,~\ref{sec:magic}).  

The sample is placed into the optical cryostat with controlled
temperature. To excite the cavity, we use a pulsed Ti:sapphire laser
producing picosecond long pulses at a repetition rate of 5~kHz. Prior
to coming into the cryostat, the pulses pass trough a long multi-mode
optical fiber and then through a monochromator. After the fiber, the
pulses have a duration of about $1$\,ns, and after the monochromator
they have spectral full width at half maximum (FWHM) of
$0.7$\,meV. The excitation is performed either along the normal to the
cavity plane ($k_\pp = 0$) or at the magic angle ($k_\pp =
1.8\,\mu\mathrm{m}^{-1}$) slightly above the lower polariton branch
resonance, with various pump polarizations (circular, linear, or
elliptical).  The main axis of pump polarization is directed along the
$\langle 110 \rangle$ axis of the structure. The pump beam is focused
onto the spot with a diameter of $100\,\mu$m.  The kinetics of the
cavity transmission signal $I_\mathrm{tr}(t)$ is detected at $T
\approx 6$\,K by the streak camera with spectral, angular, and time
resolution of $0.28$\,meV, $0.5^\circ$, and $70$\,ps, respectively, in
various polarizations.  The temporal dependences of the signal are
averaged over ${\sim}\,10^4$ pulses, for we have to collect the signal
in order to eliminate noises.

\section{Experimental results and discussion}%
\label{sec:experiment}
\subsection{Pumping at normal incidence ($k_\pp = 0$)}%
\label{sec:normal}

The information on the intra-cavity field kinetics is obtained from
the cavity transmission measurements.\cite{Demenev08,Demenev09} The
active region of the cavity is separated from a detector by the Bragg
mirror that does not introduce any nonlinearity and/or spectral
selectivity. Thus, the intensity of the transmission signal,
$\itr(\omega, t)$, is proportional to the squared magnitude of the
intra-cavity electric field $|\mathcal{E}(\mathbf{k}_\pp, \omega,
t)|^2$, and the first momentum $\bar E = \hbar \int \omega \itr(\omega, t) \,
d\omega / \int \itr(\omega, t) \, d\omega$ at $\mathbf{k} =
\mathbf{k}_\pp$ reflects the time dependence of the average
(effective) energy of excited mode $\bar{E}(t)$.

\begin{figure}[!t]
  \centering
  \includegraphics[width=0.9\linewidth]{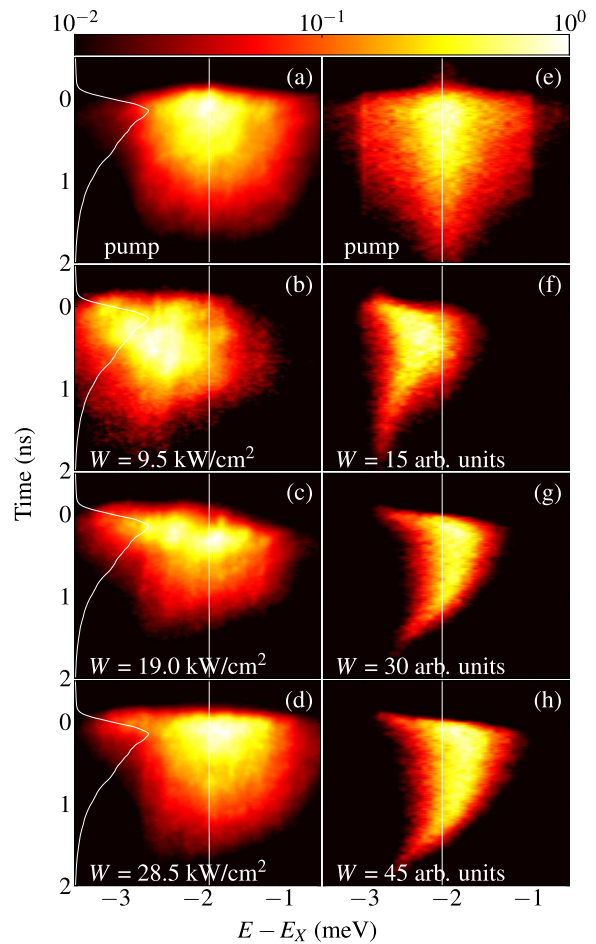}
  \caption{Time and energy dependences of the excitation pulse
    (\textit{a, e}) and cavity transmission intensity $\itr = \itr(E,
    t)$ for different pump powers $W$ (\textit{b--d} and
    \textit{f--h}). The pump is polarized circularly. The normalized
    temporal shape of the excitation is shown at the left side (white
    solid line); the weighted mean value of the pump energy is shown
    by the vertical line. Left and right panels represent the
    experimental and calculated data, respectively. The color scale is
    logarithmic. }
  \label{fig:4}
\end{figure}

Figures \ref{fig:4}\textit{b--d} show the measured time dependences of
transmission signal $\itr(t)$ under the excitation with circularly
polarized ($\sigma^+$) light. The spectra are recorded in the same
$\sigma^+$ polarization; the fraction of signal in the opposite
polarization ($\sigma^-$) does not exceed a few percent. At low peak
excitation density $W \equiv \max_t \itr(t) \approx 9.5$\,kW/cm$^2$
the maximum of the signal is found at the low energy side of the
excitation, near the lower polariton eigenfrequency. Although the
blueshift in polariton energy remains relatively small, the time
dependence of the signal differs markedly from that of the pump pulse.
This indicates the onset of nonlinear processes at the red edge of
pump spectrum. In particular, the maximum of the signal is delayed for
a few hundred picoseconds with respect to the pump
peak. Nonlinearities in both the magnitude and spectral position of
the signal increase with pump power: as the peak energy of the signal
shifts up to the pump energy $E_\mathrm{p}$, (i) the signal intensity
shows the superlinear (threshold like) increase and (ii) the time when
the signal reaches its maximum value shifts backward to the onset of
excitation pulse.

The observed correlation between the energy blueshift and intensity of
the signal is expected in the framework of the multistability
model\cite{Gippius07} Eqs.~(\ref{ef_p})--(\ref{xp_m}), however this
model fails to explain the process of instability
development. Figures~\ref{fig:4}\textit{e--h} show that the observed
time and spectral dependences can be qualitatively reproduced in
simulations using Eqs.~\eqref{ef_p_r}--\eqref{res} with appropriate
time and energy shapes of the excitation (see Appendix). The chosen
parameters of microcavity coincide with those used in the experiment;
the constants of interaction with reservoir are considered below.

\begin{figure}[!t]
  \centering
  \includegraphics[width=1.0\linewidth]{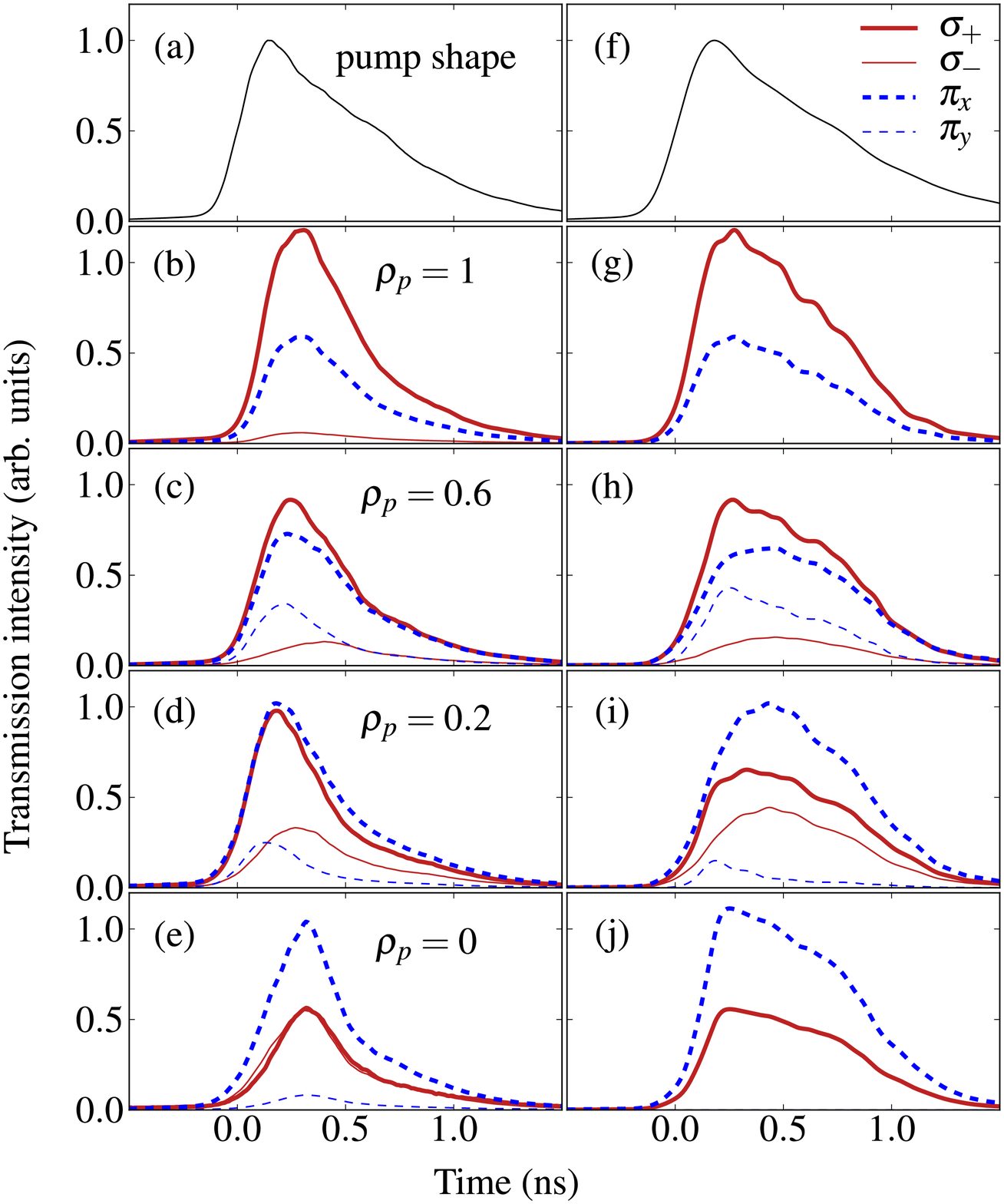}
  \caption{Time dependences of the cavity transmission intensity, in
    circular and linear polarization components ($\sigma^\pm$ and
    $\pi_{x,y}$, resp.) for different pump polarizations
    $\rho_\mathrm{p}$ (circular, elliptical, and linear
    polarizations). The top panels show the time shape of the exciting
    pulse. Left and right panels represent the experimental and
    calculated data, respectively. In the experiment, $W =
    28.5\,\mathrm{kW/cm^2}$.}
  \label{fig:5}
\end{figure}
\begin{figure}[!h]
  \centering
  \includegraphics[width=1.0\linewidth]{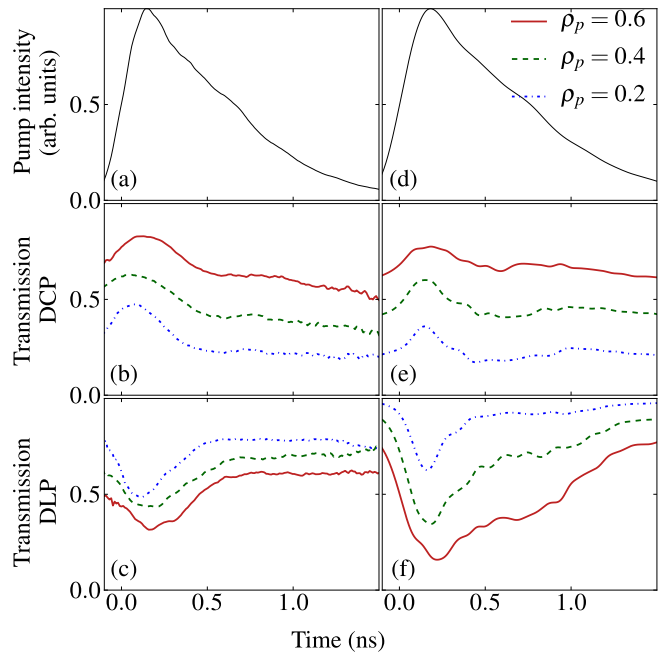}
  \caption{Time dependences of degrees of circular (\textit{b, e}) and
    linear (\textit{c, f}) polarizations of the transmitted pulses,
    for several values of the pump DCP ($\rho_\mathrm{p} = 0.2, 0.4,
    0.6$). Left and right panels represent experimental and calculated
    data, respectively. In the experiment, $W =
    28.5\,\mathrm{kW/cm^2}$.}
  \label{fig:6}
\end{figure}
\begin{figure}
  \centering
  \includegraphics[width=0.9\linewidth]{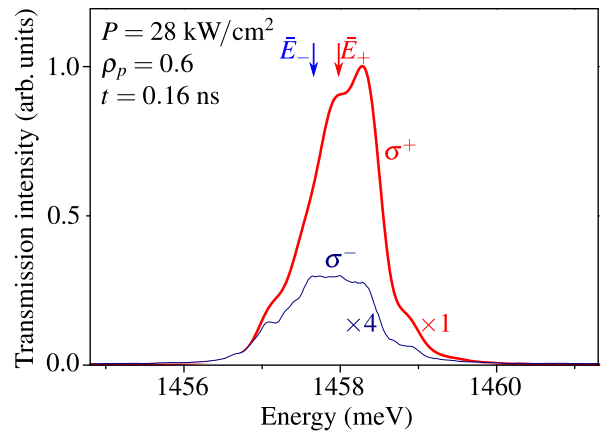}
  \caption{The spectrum of intra-cavity filed in $\sigma^\pm$
    components in the range of polarization instability
    (experiment). Weighted mean values of the energy $\bar E_\pm$ are
    indicated by arrows.}
  \label{fig:7}
\end{figure}

The temporal dependence of transmission under the excitation with
linearly, elliptically and circularly polarized pulses (DCPs 0, 0.2,
0.6, and 1) at $W = 28.5$\,kW/cm$^2$ is shown in Fig.~\ref{fig:5}. The
temporal delay of the dominating $\sigma^+$ and $\pi_x$ polarization
components of the signal with respect to the pump peak is seen to be
in the range of 150\,{--}\,200\,ps, which is much larger than the
lifetime of polaritons with $k = 0$. The ``anomalous'' increase in the
intra-cavity field observed in the range of decreasing pump power
(after the pump peak is already passed) can be explained by the
blueshift in effective polariton energy. In turn, this blueshift could
only be provided by the long-lived polariton states accumulated during
the pulse. Such an effect corresponds to the ``linear'' process of
polariton scattering into the exciton reservoir, which is determined
by the ratio between $\gamma_\mathrm{r}$ and $\gamma_\mathrm{xr}$ in
Eqs.~(\ref{xp_p_r})--(\ref{res}). The experimental results are
qualitatively reproduced in numerical simulations with $V_2 / V_1 =
-0.1$, $V_\mathrm{r} / V_1 = 7 \cdot 10^{-3}$, $\hbar
\gamma_\mathrm{r} = 2 \cdot 10^{-3}$\,meV, and $\gamma_\mathrm{xr} = 3
\cdot 10^{-3}$\,meV. These parameters are chosen to meet the following
conditions: (i) the overall occupation of the reservoir is comparable
to that of the driven (optically active) polariton mode, so that
$W_\mathrm{thr}^\mathrm{(lin)} \lesssim
W_\mathrm{thr}^\mathrm{(circ)}$, and (ii) the temporal peak of the
signal can be delayed by hundreds of picoseconds with respect to the
excitation peak even for $\rho_\mathrm{p} = \pm 1$. Note the long
decay times of the reservoir states, ${\sim}\,300$\,ps. So long times
are characteristic of excitons localized due to fluctuations of
quantum well potential and/or free excitons with large lateral wave
numbers.

Figures \ref{fig:5}\textit{b} and \ref{fig:5}\textit{e} reveal the
transmission signal to be retaining both circular and linear pump
polarizations; in these cases only a weak depolarization of the
transmitted light is observed. Retaining of the circular polarization
in agreement with the angular momentum conservation law is
expected. Measurements performed under linearly polarized pump in a
wide range of $W = 9.5\,{-}\,28.5$\,kW/cm$^2$ (not shown) have
revealed the magnitude of transmitted light depolarization to be
within the limits of 12--17\%, almost independently of the pump
power. On the other hand, under the elliptically polarized excitation
($0 < \rho_\mathrm{p} < 1$) the signal DCP does depend on the pump
intensity due to the anisotropy in polariton-polariton
interaction.\cite{Gippius07} Accordingly, the ratio between both
circularly and linearly polarized components of the signal intensity
($\sigma^\pm$ and $\pi_{x,y}$, respectively) varies with time
(Fig.~\ref{fig:5}\textit{c,d}). The corresponding temporal dependences
of the degrees of circular and linear polarizations of the signal are
shown in Fig.~\ref{fig:6}. From comparison of Figs.~\ref{fig:5} and
\ref{fig:6}, it is seen that the reaching of maximum signal intensity
at $t \approx 0.2\,\mathrm{ns}$ is accompanied by a decrease in the
signal DCP, in much the same way as the cw driven system behaves
(Fig.~\ref{fig:1}), though the latter system exhibits the expectedly
sharper transitions. On the other hand, our experimental technique
allows us to visualize the energies of intra-cavity field in addition
to their intensities. The information on the energies is extracted
from the transmission spectra as shown in Fig.~\ref{fig:7} since
$I_\mathrm{tr}(\hbar\omega) \propto |\mathcal E(\hbar\omega)|^2$.  The
energies of intra-cavity field in $\sigma^+$ and $\sigma^-$
polarizations demonstrate a well pronounced blueshift with increasing
polariton density and a well pronounced splitting. In particular it is
seen in Fig.~\ref{fig:7} that in the range of maximum polarization the
$\sigma^+$ component of the field is markedly (by 0.3 meV)
blue-shifted with respect to $\sigma^-$.

The basis in which shown degree of linear polarization (DLP) was
measured is related to the pump polarization axis. We observe no
linear polarization in the basis rotated by $45^\circ$ with respect to
the former one. Thus, the sum of squared Stokes parameters per signal
intensity is less than 1, which means that the signal is partially
depolarized; this is because the signal polarization components are
measured as averaged quantities.

\begin{figure}[!t]
  \centering
  \includegraphics[width=1.0\linewidth]{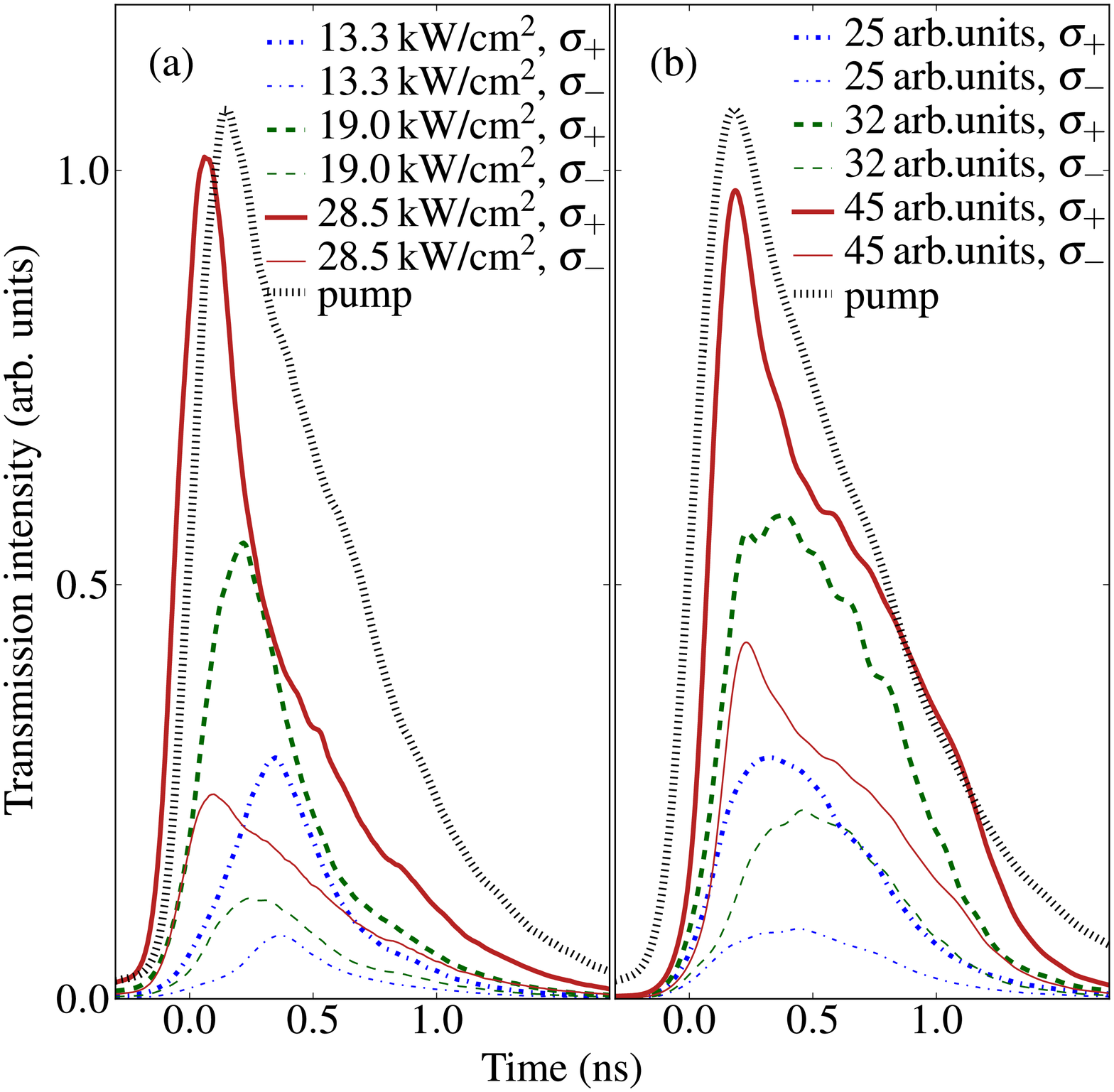}
  \caption{Time dependences of the transmission DCPs, for several
    values of the pump power $W$, under excitation with elliptically
    polarized pulses with $\rho_\mathrm{p} = 0.4$. Grey dotted line
    corresponds to the normalized temporal shape of excitation. Left
    and right panels represent experimental and calculated data,
    respectively.}
  \label{fig:8}
\end{figure}

\begin{figure}[!h]
  \centering
  \includegraphics[width=1.0\linewidth]{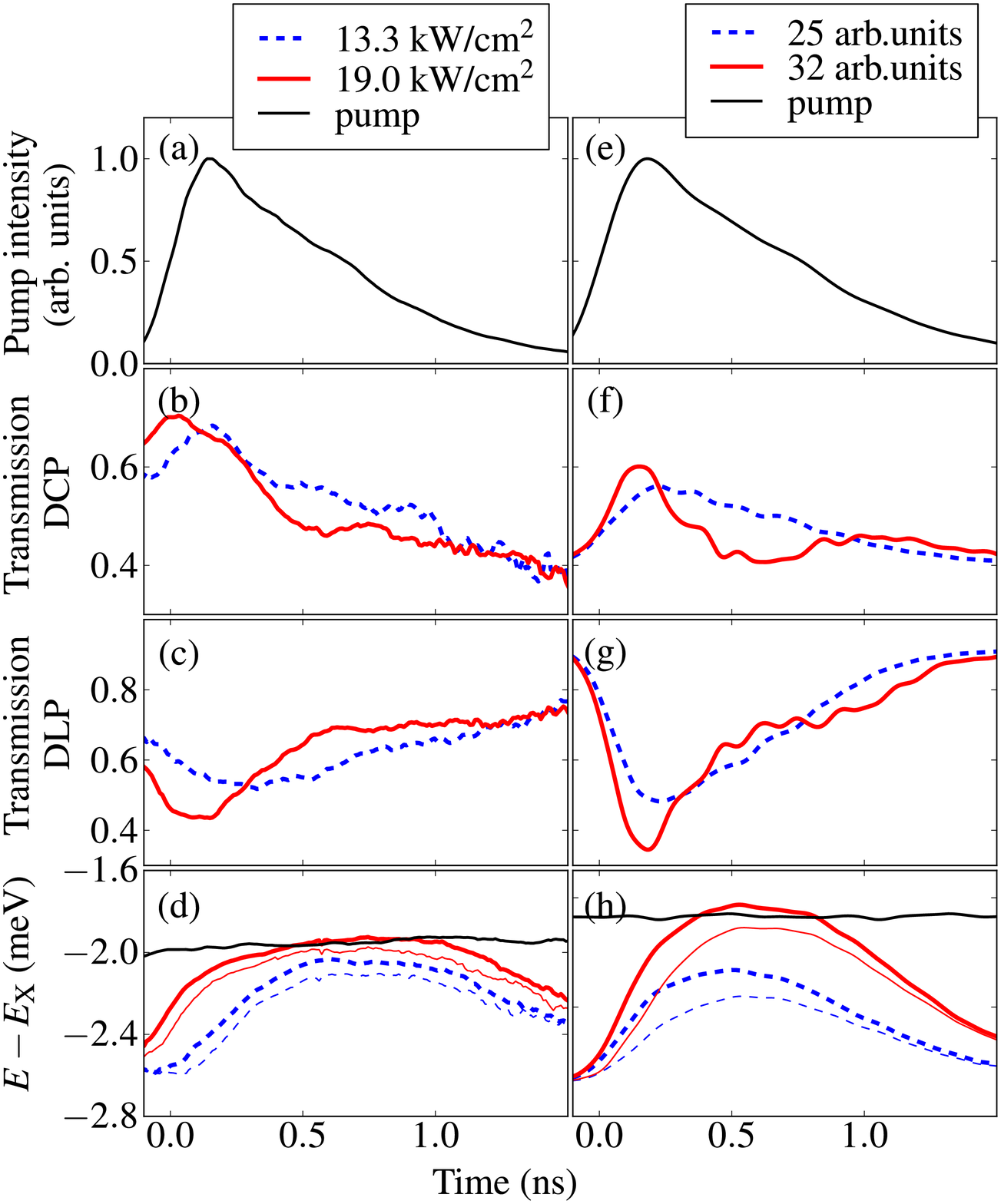}
  \caption{Time dependences of the pump pulse intensity (\textit{a,
      e}), degrees of circular (\textit{b, f}) and linear (\textit{c,
      g}) polarizations (DCP and DLP) of the transmitted pulse, and
    the weighted mean values $\bar{E}_\pm$ of the transmission energy
    in $\sigma^+$ (thick lines) and $\sigma^-$ (thin lines)
    polarization components (\textit{d, h}), for several values of the
    pump power $W$. The pump DCP is $0.4$. The mean value of pump
    energy, which exhibits minor variations in a close vicinity of $E
    - E_\mathrm{x} = -2\,\mathrm{meV}$, is shown by solid thin lines
    in panels (\textit{d, h}). Left and right panels represent
    experimental and calculated data, respectively.}
  \label{fig:9}
\end{figure}
\begin{figure}
  \centering
  \includegraphics[width=1\linewidth]{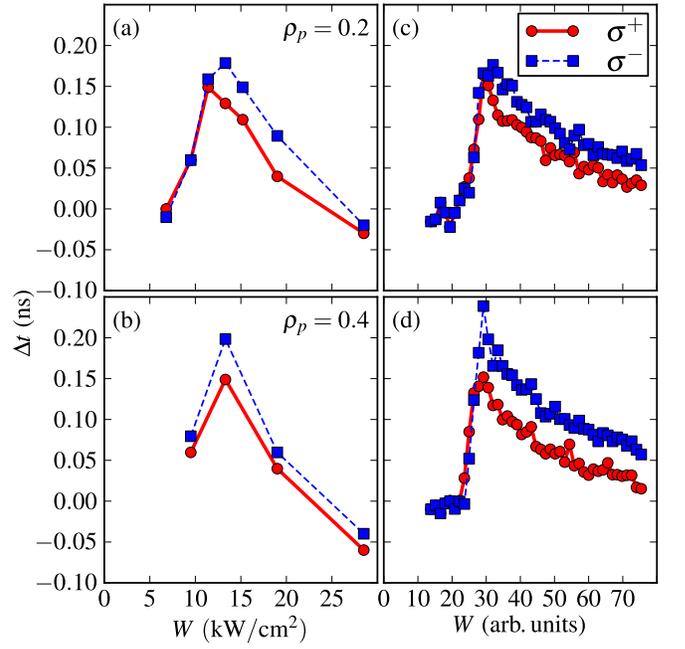}
  \caption{The effective delay of the $\sigma^\pm$ components of the
    signal with respect to the pump vs. pump power (as defined by
    Eq.~(\ref{eq:delay})), for two pump polarizations, $\rho_p = 0.2$
    (panels \textit{a, c}) and $\rho_p = 0.4$ (panels \textit{b,
      d}). Left and right sides of the figure represent experimental
    and calculated data, respectively.}
  \label{fig:10}
\end{figure}

In order to clarify the physics underlying the signal polarization
transitions, let us consider the series of measurements performed
under elliptically polarized pulses with $\rho_\mathrm{p} = 0.4$ and
various pump intensities.  Fig.~\ref{fig:8} represents measured (left
side) and calculated (right side) time dependences of the transmission
signal in $\sigma^\pm$ polarization components. From these series, the
pump power dependence of the temporal delay between the signal and
pump peaks is evidenced.  The certain conditions for a switch in
signal polarization depend on both the pump intensity and its temporal
shape. In case of the very low pump densities, the intra-cavity field
is insufficient for multistability to reveal; on the other hand, if
the pump density is increasing quickly enough then a jump into a
high-energy state happens with no delay related to the long-lived
reservoir, since most of the intra-cavity field is concentrated in the
driven (optically active) polariton state. Except the case of maximum
$W$\!, the dynamics shown in Fig.~\ref{fig:8} lies in the intermediate
range of pump powers which all are insufficient to pull a polariton
state by itself up to the threshold magnitude. Instead of that, the
non-equilibrium transitions proceed through the mediation of
excitation accumulated by the reservoir. In such a case the
development of instability in the leading polarization component
($\sigma^+$) always precedes the reservoir induced switch in signal
polarization.

Figure~\ref{fig:9}, which also shows a series for two peak pump
intensities at fixed $\rho_\mathrm{p} = 0.4$, allows us to visualize
the temporal correlations between the energy (in both $\sigma^\pm$
polarization components) and polarization degree of the transmission
signal. The overall process of the instability development is
accompanied by blueshifts in $\sigma^\pm$ energies. At the first stage
(which starts the earlier, the higher is pump density $W$), the
leading polarization component ($\sigma^+$) enters the region of
positive feedback between its effective resonance energy $\bar{E}_+$
and the $\sigma^+$ field amplitude, so the difference in $\sigma^\pm$
energies grows along with the DCP of transmission signal
(Fig.~\ref{fig:9}\textit{b, d}). Accordingly, a well pronounced
decrease in the DLP of transmission is observed at the same time
interval (Fig.~\ref{fig:9}\textit{c}). At approximately the half way
through the energy growth (see panel~\textit{d}), the integral
occupation of reservoir becomes sufficient to provide the shift in the
minor component ($\sigma^-$), which leads to decrease in both the
energy difference $\bar{E}_+ - \bar{E}_-$ and DCP, and to the
corresponding increase in DLP. All these effects are qualitatively
reproduced in calculations shown in
Fig.~\ref{fig:9}\textit{e--h}. Thus, the observed dynamics confirm the
effect of reservoir, though the energy broadening makes the
transitions much smoother compared to the cw driven system
(Fig.~\ref{fig:1}).

Figure~\ref{fig:10} shows the time delay of the signal with respect
to the excitation pulse in a wider range of pump powers $W$\!; it
summarizes the consideration of the threshold effects in the cavity
transmission. The value of the delay $\Delta t$ is determined as the
difference between the moments the signal and the pump reach their
half-maxima for the first time:
\begin{equation}
  \Delta t_\pm = \displaystyle t \left|_{I_\mathrm{tr}^\pm(t) \to \frac12 \max_t
      I_\mathrm{tr}^\pm(t)} \right. - t \left|_{I_\mathrm{p}(t) \to \frac12
      \max_t I_\mathrm{p}(t)} \right..
  \label{eq:delay}
\end{equation}
At small $W$ the response is linear, and the delays $\Delta t$ are
expectedly small. At threshold, the superlinear growth of the signal
is observed with $\Delta t$ of about 150 to 200 picoseconds, i.\,e.\
at the back front of excitation. Thus, this transition is induced by
the reservoir as explained above. With further increasing $W$, the
optically active fraction of excitons grows, which lowers the system's
intertness, and the delay of response decreases. Note, that the minor
($\sigma^-$) polarization component exhibits the superlinear growth at
the same \emph{peak} power $W$ as $\sigma^+$, and the relative delay
of $\sigma^+$ and $\sigma^-$ components, $\Delta t_+ - \Delta t_-$,
remains nearly constant above the threshold. This is an extra evidence
of the fact that the jump in the minor component is provided by the
reservoir even at high $W$\!.

\subsection{Pumping at the magic angle ($k_\pp =
  1.8\,\mu\mathrm{m}^{-1}$)}%
\label{sec:magic}

Let us now consider the polarization properties of the signal of
parametric scattering---generally referred to as cavity ``optic
parametric oscillator'' (OPO) signal\cite{Kavokin-book-03}---which
appears at the branch bottom ($k = 0$) if pumping is at the magic
angle ($k_\pp \approx 1.8\,\mu\mathrm{m}^{-1}$). There are two points
to discuss here. First, in the general case of elliptically polarized
excitation the signal is expected to reflect variations in the DCP of
the driven mode that, in turn, are shown to be strongly affected by
the excitonic reservoir. Second, if the pump is polarized linearly,
the experiments show the $90^\circ$ rotation of the signal
polarization axis;\cite{Krizhanovskii06-prb} thus, we have to check if
Eqs.~(\ref{ef_p_r})--(\ref{res}) can reproduce this effect in spite of
the reservoir induced interaction between the cross-circularly
polarized excitons.

The experiments are carried out at negative detuning between the
photon and exciton ($\Delta \approx -1.5\,\mathrm{meV}$) in order to
have the pump frequency close to that employed for the normal
incidence pumping considered in Sec\,~\ref{sec:normal}. The cavity and
excitation parameters used in the simulations coincide with the
experimental ones, whereas the nonlinear interaction constants as well
as reservoir characteristics are the same as used in the previous
Section.

\begin{figure}[!t]
  \centering
  \includegraphics[width=1.0\linewidth]{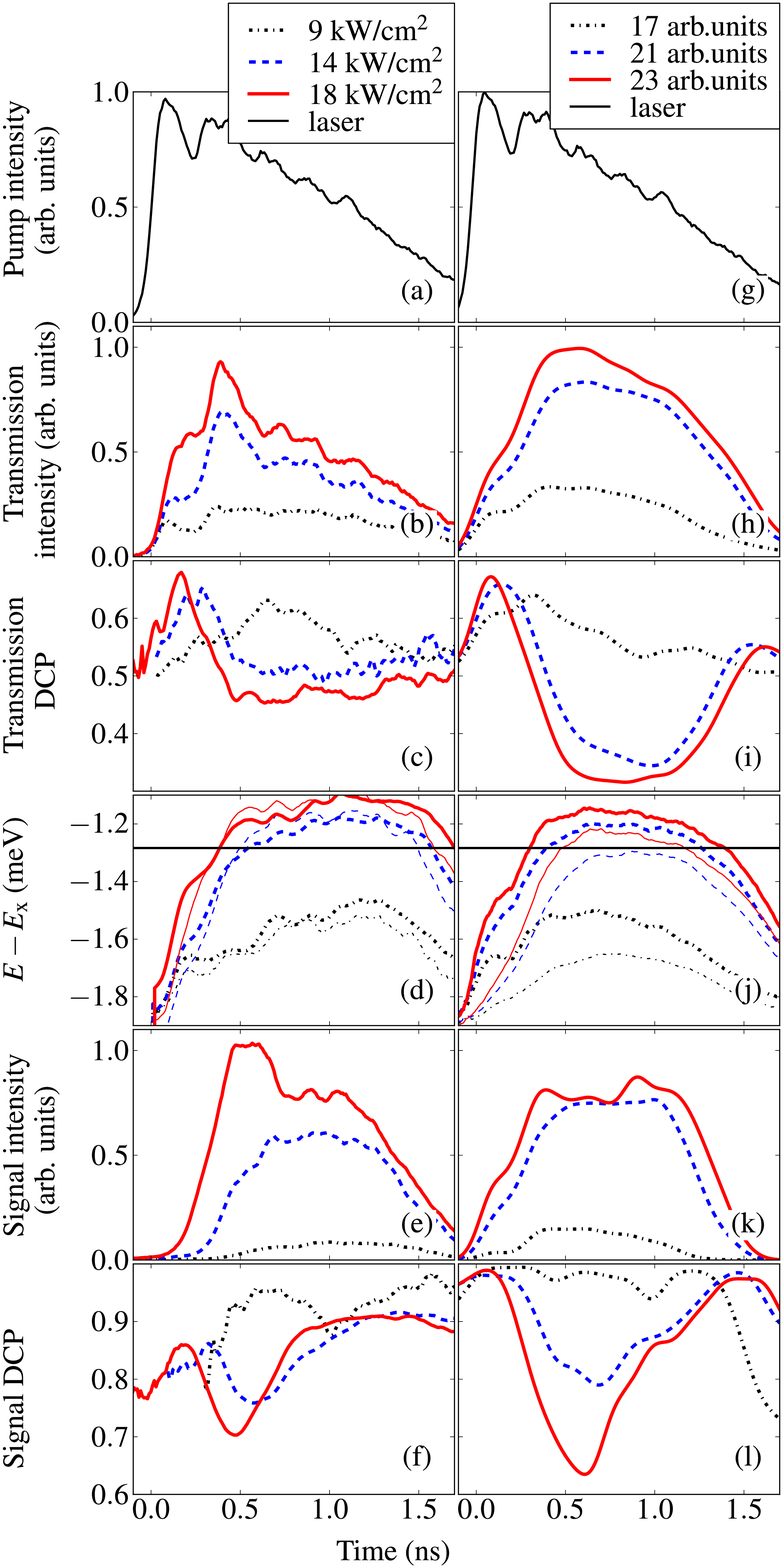}
  \caption{Time dependences of the pump pulse intensity (\textit{a,
      g}), transmission intensity (\textit{b, h}), transmission DCP
    (\textit{c, i}), weighted mean values of the transmission energy
    $\bar{E}_\pm$ (\textit{d, j}) in two polarization components
    $\sigma^+$ and $\sigma^-$ (thick and thin lines, resp.), OPO
    signal intensity (\textit{e, k}) and DCP (\textit{f, l}), for
    several values of the pump power $W$. Straight lines in panels
    \textit{c, i} indicate the mean value of pump energy. Left and
    right panels represent experimental and calculated data,
    respectively.}
  \label{fig:11}
\end{figure}
\begin{figure}[!t]
  \centering
  \includegraphics[width=1.0\linewidth]{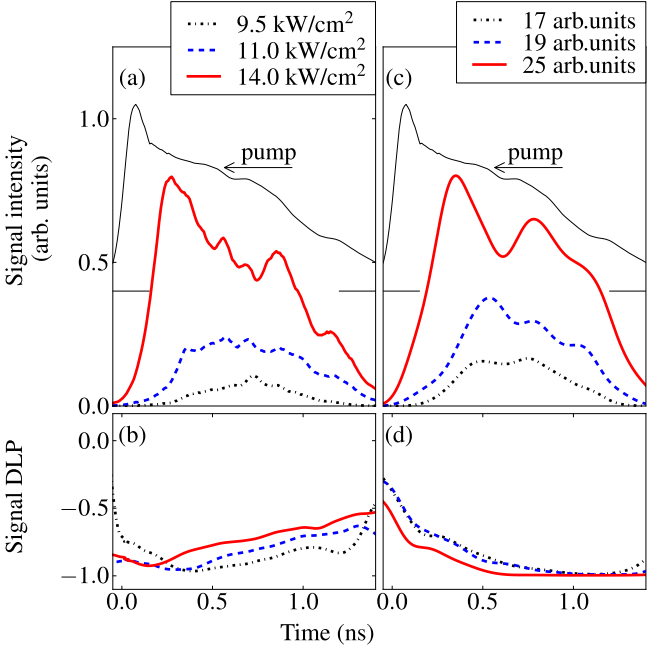}
  \caption{Time dependences of the OPO signal intensity (\textit{a,
      c}) and degree of linear polarization (\textit{b, d}) under the
    pump $\vec{\mathcal{F}}(t)$ polarized linearly along the TM cavity
    mode ($\vec{\mathcal{F}} \parallel
    \vec{\mathcal{E}}_\mathrm{TM}$), for some different values of the
    pump power $W$. Left and right panels represent the experimental
    and calculated data, respectively. The temporal shape of
    excitation is indicated in panels \textit{a, c} as a guide for
    eye.}
  \label{fig:12}
\end{figure}
\begin{figure}[!t]
  \centering
  \includegraphics[width=1.0\linewidth]{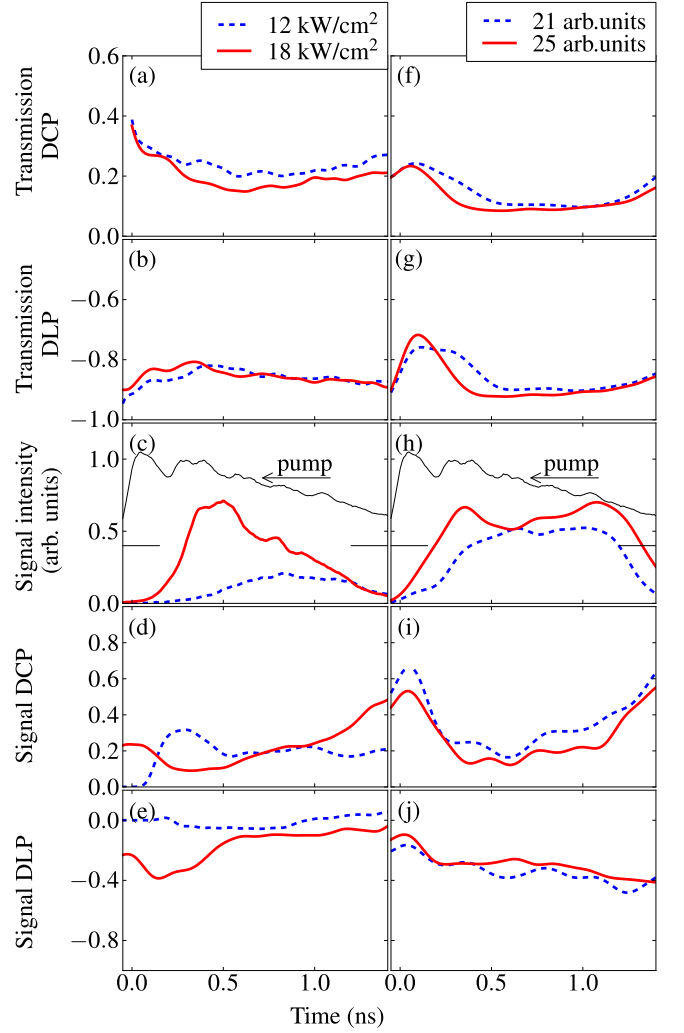}
  \caption{Time dependences of the transmission DCP (\textit{a, f}),
    transmission DLP (\textit{b, g}), signal intensity (\textit{c,
      h}), signal DCP (\textit{d, i}), and signal DLP (\textit{e, j})
    under linearly polarized pump wave $\vec{\mathcal{F}} \parallel
    (\vec{\mathcal{E}}_\mathrm{TE} +
    \vec{\mathcal{E}}_\mathrm{TM})$. Left and right panels represent
    experimental and calculated data, respectively. The temporal
    shapes of excitation are indicated in panels \textit{a, c} as a
    guide for eye. In calculation, the TE/TM splitting of the driven
    polariton mode equals to $0.06$\,meV.}
  \label{fig:13}
\end{figure}

Figure~\ref{fig:11} represents measured and calculated dynamics of
intra-cavity field at $k = k_\mathrm{p}$ (which corresponds to the
cavity transmission) and $k = 0$ (OPO signal) under elliptically
polarized excitation with $\text{DCP} = 0.5$ and different pump
powers~$W$\!. When $W$ exceeds the critical value (in the experiment
$W_\mathrm{thr} \approx 10\,\mathrm{kW/cm^2}$), a well pronounced
superlinear growth of the cavity transmission
(Fig.~\ref{fig:11}\textit{b}) is observed at $t =
0.1\,{\text{--}}\,0.4\,\mathrm{ns}$. Near the threshold, a doubling of
the pump power~$W$ leads to an order-of-magnitude growth of the OPO
signal intensity (Fig.~\ref{fig:11}\textit{e}). The observed
nonlinearities indicate the onset of parametric instability, being,
however, by far not as sharp as under cw
excitation.\cite{Krizhanovskii08}

The dynamics of the driven mode is similar to that under the pumping
at normal incidence.  The growth of the field amplitude, which
continues with increasing pump density (Fig.~\ref{fig:11}\textit{b}),
expectedly leads to the growth of transmission DCP
(Fig.~\ref{fig:11}\textit{c}) due to the dominantly repulsive
exciton-exciton interaction. At $t = 0.1\,{\text{--}}\,0.4$\,ns the
transmission starts to behave strongly nonlinearly with respect to the
driving field, and the peak of transmission becomes significantly
delayed with respect to the peak of pump pulse. The latter indicates
that the transition is assisted by the reservoir induced
blueshift. Accordingly, the jump in transmission DCP is immediately
followed by dropping back as soon as the system enters the
above-threshold region where the energies of $\sigma^\pm$ components
of intra-cavity field (Fig.~\ref{fig:11}\textit{d}) become partially
levelled due to the reservoir filling.

The domination of the leading polarization component ($\sigma^+$) is
strongly enhanced in the OPO signal (Fig.~\ref{fig:11}\textrm{f}) whose
DCP reaches~0.85 at $t \approx 0.2$\,ns. The subsequent lowering of
DCP of the driven mode forces the OPO signal, too, to partially loss
its polarization. However, the signal restores a high DCP during the
further evolution accompanied by the decrease in transmission
intensity. The calculations performed on the basis of
Eqs.~(\ref{ef_p_r})--(\ref{res}), which are presented in the right
panels of Fig.~\ref{fig:11}, reproduce the observed threshold-like
behavior of intra-cavity field along with its polarization
properties. However, a quantitative agreement is not achieved,
partially due to a complicated spectral shape of the pump pulses (see
Fig.~\ref{fig:4}) and due to intrinsic limitations of the suggested
theoretical model (see discussion in Sec.~\ref{sec:discussion}).

Let us turn to the case of exactly linear polarization of the
excitation at $k_\pp > 0$ (Figs.~\ref{fig:12} and \ref{fig:13}). In
this case the OPO signal depends on whether the pump excites a pure
cavity state---one of TE and TM modes which are the eigenstates of 2D
photon in the empty cavity---or a mixture of both the TE and TM
components. Fig.~\ref{fig:12} represents the dynamics of OPO signal
under the pump polarized along the TM cavity mode. In the range of
high intensity of the signal, its degree of linear polarization (DLP)
reaches 90\%. The polarization axis is $90^\circ$ rotated with respect
to the pump (what corresponds to negative DLP values), which is well
reproduced in the calculations.

On the other hand, if the pump is polarized in between the TE and TM
directions ($\vec{\mathcal{F}} \parallel
[\vec{\mathcal{E}}_\mathrm{TE} + \vec{\mathcal{E}}_\mathrm{TM}]$,
Fig.~\ref{fig:13}), the signal DLP does not exceed 30\% in the range
of a strong signal (Fig.~\ref{fig:13}\textit{e}). The shown DLP values
were measured (and calculated) in the basis of directions parallel and
orthogonal to pump polarization; there is no marked DLP in the TE-TM
basis as well.  This effect is explained by the lifted degeneracy
(TE-TM splitting) of cavity modes which leads to the variation with
time of the $\sigma^\pm$ phase shift. Accordingly, it causes the
misbalance of $\sigma^+$ and $\sigma^-$ components of intra-cavity
field, i.\,e.\ a reduced DLP and non-zero DCP of the driven mode
(Fig.~\ref{fig:13}\textit{a, b}), which is seen in both measured and
calculated data. Due to permanent modification of the direction of
signal polarization, the eventual OPO signal appears to be
significantly depolarized (the sum of squared DCP and DLP is markedly
less than~1, Fig.~\ref{fig:13}\textit{d, e}).

Thus, the dynamics of the OPO signal confirms the effect of
reservoir. Under elliptic excitation, the $\sigma^\pm$ polarization
components tend to be levelled by the reservoir. By itself, this
effect could be assigned to a repulsion between cross-circularly
polarized excitons ($V_2 > 0$). On the other hand, the axis of signal
polarization is found inverted with respect to the pump, which
restricts the sign of $V_2$ to be negative irrespectively of the
reservoir induced blueshifts. Since we have reproduced these effects
within Eqs.~(\ref{ef_p_r})--(\ref{res}), it is proven to provide a
self-consistent description of the intra-cavity field polarization
dynamics.

\subsection{Shortcomings}%
\label{sec:discussion}

We have shown that the semi-phenomenological
Eqs.~(\ref{ef_p_r})--(\ref{res}) explain the observed polarization
dynamics in a good qualitative agreement with the experimental data.
Let us now consider the limitations of the suggested approach. Its
most important shortcoming is the absence of a feedback between the
driven polariton modes and the exciton reservoir, as long as the
latter is introduced using the integral occupation
number~$\mathcal{N}$, neglecting the actual microscopic distribution
of reservoir states. As a result, Eqs.~(\ref{ef_p_r})--(\ref{res}) do
not describe, e.\,g.\, the establishing of equilibrium between
coherent and incoherent exciton ``phases'' which would normally take
place in a real system; this process implies (i) the reverse
transitions from the reservoir into the macro-occupied mode (similar
to Bose-Einstein condensation) and (ii) dependence of the scattering
rates ($\gamma_\mathrm{xr}$ and $V_\mathrm{r}$) on the polariton
density. These shortcomings lead to several visible discrepancies
between the measured and calculated data. In particular, the
superlinear drop in transmission intensity observed at the back front
of exciting pulses, which is evident from comparison of left and right
panels of Fig.~\ref{fig:8}, is not reproduced by the present
model. Note as well that the nonlinear scattering rate $V_\mathrm{r}$
should depend on the energy of biexciton resonance which mediates the
transitions of optically driven polaritons into the
reservoir.\cite{Wouters07,Schumacher07} Accordingly, the actual
relation between the multistability thresholds
$W_\mathrm{thr}^\mathrm{(circ)}$ and $W_\mathrm{thr}^\mathrm{(lin)}$
would depend on the pump frequency as well as on the frequency
detuning between photon and exciton modes.

As a result, we expect Eqs.~(\ref{ef_p_r})--(\ref{res}) to be
qualitatively valid in the case of resonant excitation of the
optically active cavity states. On the other hand, in the cases of
comparably large pump wave numbers and/or non-resonant excitation
conditions, when a much larger contribution of reservoir excitons is
expected, the model is likely to become unsatisfactory.

\section{Conclusion}%
\label{sec:conclusions}

In the present work we have studied the non-equilibrium transitions in
a multistable system of cavity polaritons under resonant
nanosecond-long excitation. Using the spectrally broadened pulses, we
have visualized the temporal correlations between the effective
resonant energy, intensity, and optical polarization of the
intra-cavity field which all undergo the strong changes on reaching
the threshold pump power. In the vicinity of the threshold, the
dynamics of such system is strongly affected by the long-lived exciton
reservoir (excited due to polariton scattering) which influences both
the characteristic times of instability development and output signal
polarization.

The temporal behavior of the intra-cavity field is found to be not
described in the conventional model based on the Gross-Pitaevskii
equations written for purely coherent macro-occupied polariton
states. Most importantly, the observed phenomena cannot be explained
even qualitatively within a model with only the two exciton-exciton
interaction constants ($V_{1,2}$) allowed for. To explain the
experiments, we have proposed the model for the macro-occupied
polariton states coupled with an exciton reservoir. In spite of some
limitations, this model provides a self-consistent description of the
observed intra-cavity field dynamics under both pulse and continuous
wave excitation conditions.

Authors thank M.\,S.\,Skolnick for rendered samples,
D.\,N.\,Krizhanovskii for fruitful discussions, and A.\,V.\,Larionov
for assistance in the experiment. This work was supported by 
Russian Foundation for Basic Research and the Russian Academy of Sciences.

\appendix*

\section{Details on numeric modeling}
\label{sec:modeling}

In attempt to reproduce the experimental pump spectrum, which is
characterized by FWHM of about 0.8\,meV and, hence, has a lowered
coherence, the modeled pump is `broadened' by means of the partial
randomization of phases of different spectral harmonics. Namely, the
pump amplitude is defined as follows:
  \begin{gather}
    \mathcal F_\pm^{\vphantom )}(t) = \mathcal F_\pm^{(0)}(t) \left(
      \frac{\mathcal R(t)}{|\mathcal R(t)|} \right) \exp \left(
      -i\omega_p
      t - i\varphi_\pm \right) ; \\
    \mathcal R(t) = \sum_{n=1}^N \exp \left(
      -\frac{\omega_n^2}{2\sigma^2} -i \omega_nt - \frac{2i\pi}{N}
      \sum_{m=1}^n \chi_m \right).\label{eq:pump}
  \end{gather}
Here, 
\begin{itemize}
\item $\mathcal F_\pm^{(0)}(t)$ are real-valued amplitudes of the
  $\sigma^\pm$ polarization components; 
\item $\varphi_+ - \varphi_-$ defines the direction of the
  polarization axis;
\item $\omega_p$ is the central pump frequency;
\item $\omega_n = \Omega \left( \frac{n}{N} - \frac12
  \right)\!, ~ n = 1,2, \ldots N; ~ N=1000; ~ \Omega =
  2\,\mathrm{meV}$;
\item $\sigma = \mathrm{FWHM}/\sqrt{8 \log 2}$ defines the width of
  the pump spectrum, $\mathrm{FWHM} = 0.8\,\mathrm{meV}$;
\item $\{\chi_m \mid m = 1,2,\ldots,N\}$ is a set of random uniformly
  distributed numbers within the interval $[0;1)$.
\end{itemize}%
The spectral FWHM of the pump defined such a way approximately
coincides with experimental one, and its coherence time is about 5
picoseconds. Under these conditions the system is inevitably
stochastic. Hence, for each set of parameters
Eqs.~(\ref{ef_p_r})--(\ref{res}) are solved a number of times
(20--100) with different $\{\chi\}$ until the averaged evolution of
the signal intensity becomes independent on the number of
realizations. This corresponds to the collection of numerous signals,
which allows us to reduce noises.  The degrees of polarization are
then calculated using the \emph{averaged} intensities. As a result,
the transmission can be partially depolarized (under elliptically
polarized excitation) due to small random differences in the
individual pulses in much the same way as the measured signal.


%

\end{document}